# Multi-stage Resilience Management of Smart Power Distribution Systems: A Stochastic Robust Optimization Model

Nariman L. Dehghani, *Student Member, IEEE,* Abdollah Shafieezadeh, *Member, IEEE*

*Abstract*—Significant outages from weather and climate extremes have highlighted the critical need for resilience-centered risk management of the grid. This paper proposes a multi-stage stochastic robust optimization (SRO) model that advances the existing planning frameworks on two main fronts. First, it captures interactions of operational measures with hardening decisions. Second, it properly treats the multitude of uncertainties in planning. The SRO model coordinates hardening and system operational measures for smart power distribution systems equipped with distributed generation units and switches. To capture the uncertainty in the incurred damage by extreme events, an uncertainty set is developed by integrating probabilistic information of hurricanes with the performance of overhead structures. A novel probabilistic model for the repair time of damaged lines is derived to account for the uncertainty in the recovery process. A solution strategy based on the integration of a differential evolution algorithm and a mixed-integer solver is designed to solve the resilience maximization model. The proposed approach is applied to a modified IEEE 33-bus system with 485 utility poles and a 118-bus system with 1841 poles. The systems are mapped on the Harris County, TX, U.S. Results reveal that optimal hardening decisions can be significantly influenced by resilience operational measures.

## NOMENCLATURE

*Variables:*

$\mathcal{C}^h, \mathcal{C}^d, \mathcal{C}^{ld}$, and $\mathcal{C}^r$: Cost of hardening, damage, load shedding and recovery

$x$: Discrete pole hardening decision variables; $x_{ij}$ indicates total number of replaced poles in line $(i,j)$

$u$: Binary line damage variables; $u_{ij,t} = 1$ if line $(i,j)$ at time $t$ is damaged, and 0 otherwise

$z$: Power flow variables, *i.e.,* $P_{ij,t}, Q_{ij,t}, V_{j,t}, P_{j,t}^g, Q_{j,t}^g$, and $\rho_{j,t}^{ld}$

$s$: Binary switch variables; $s_{ij,t} = 1$ if switch at line $(i,j)$ is open at time $t$, and 0 otherwise

$y$: Binary status variables; $y_{ij,t} = 1$ if line $(i,j)$ is disconnected at time $t$ (i.e., if $u_{ij,t} = 1$ and/or $s_{ij,t} = 1$), and 0 otherwise

$r$: Discrete recovery decision variables; $r_{ij,t}$ indicates number of repair crews repairing line $(i,j)$ at time $t$

$\delta$: Binary travel variables; $\delta_{ij,t} = 1$ if any repair crew traveled to line $(i,j)$ at time $t$, and 0 otherwise

This research was partly funded by the U.S. National Science Foundation (NSF) through award CMMI-2000156. This support is greatly appreciated.

N. L. Dehghani is with the Risk Assessment and Management of Structural and Infrastructure Systems lab, Ohio State University, Columbus, OH 43210 USA (e-mail: laaldehghani.1@osu.edu).

A. Shafieezadeh is with the Risk Assessment and Management of Structural and Infrastructure Systems lab, Ohio State University, Columbus, OH 43210 USA (e-mail: shafieezadeh.1@osu.edu).

$\boldsymbol{\beta}$: Binary spanning tree variables; $\beta_{ij,t} = 1$ if bus $i$ is the parent node to bus $j$ at time $t$, and 0 otherwise

$\boldsymbol{\xi}$: Weather related uncertain variables; *i.e.,* hurricane wind speed and wind direction

*$\boldsymbol{\mu, \lambda}$:* Dual variables

$P_{ij,t}/Q_{ij,t}$: Active/reactive power flow in line $(i,j)$ at time $t$

$V_{j,t}$: Voltage at bus $j$ and time $t$

$P_{j,t}^g/Q_{j,t}^g$: Active/reactive power generated at bus $j$ and time $t$

$\rho_{j,t}^{ld}$: Ratio of load shedding at bus $j$ and time $t$

$P_{ij}^f$: Expected annual failure probability of line $(i,j)$

$p_{m,ij}^f$: Expected annual failure probability of pole $m$ in line $(i,j)$

$r_{ij}$: Uncertain repair time of line $(i,j)$

$N_{ij}^f$: Uncertain number of pole failures in line $(i,j)$

*Sets:*

$\mathbb{X}$: Feasible set of hardening decisions

$\mathbb{U}$: Uncertainty set of damages

$\mathbb{O}$: Feasible set of system operation

$\mathbb{R}$: Feasible set of repair decisions

$\psi_{ij}$: Set of utility poles in line $(i,j)$

$\phi(j)$: Parent node of bus $j$

$\varphi(j)$: Set of child nodes of bus $j$

$\Omega_L$: Set of distribution lines

$\Omega_B$: Set of buses

$\Omega_{RB}$: Set of root buses

$\Omega_{SW}$: Set of lines with switches

$\Omega_F$: Set of failed lines

$\Omega_T$: Set of time of interests

$P_g^p$: Parent population in generation $g$

$Q_g^p$: Offspring population in generation $g$

$H_g^p$: Combined population in generation $g$

*Parameters:*

$B_H$ and $B_U$: Hardening budget and uncertainty budget

$P_{j,t}^{ld}/Q_{j,t}^{ld}$: Active/reactive load demand at bus $j$ and time $t$

$P_{ij}^{max}/Q_{ij}^{max}$: Maximum active/reactive flow in line $(i,j)$

$P_j^{g,max}/Q_j^{g,max}$: Max active/reactive power generated at bus $j$

$|V_j|^{min}/|V_j|^{max}$: Min/max voltage magnitude at bus $j$

$R_{ij}$ and $X_{ij}$: Resistance and reactance of line $(i,j)$

$V_0$: Reference voltage

$M_1, M_2$, and $M_3$: Big numbers

$\varepsilon_1$ and $\varepsilon_2$: Small numbers

$n$: Total number of repair crews

$h^r$: Required hours for repairing a pole by a repair crew

$t_e$: Time of event occurrence

$t_r$: Start time of recovery process

$t_c$: Control time

$c^h$: Cost coefficient of replacing a utility pole

$c_{j,t}^{ld}$: Cost coefficient of load shedding at bus $j$ and time $t$

$c_{ij,t}^{rp}$: Cost coefficient of a crew repairing line $(i,j)$ at time $t$

$c^{tr}$: Cost coefficient of traveling between two damaged lines

$n_g$: Total number of generations

$n_p$: Number of solutions in parent population

$F_s$ and $C_r$: Scale factor and crossover rate

# I. INTRODUCTION

Extreme weather and climate events have been a significant threat to built, natural, and human systems across the world. As illustrated in Fig. 1, the number of severe storms and tropical cyclones – also referred to as hurricanes – and the losses imparted by these events have been rising in the U.S. over the past 40 years [1]. This trend is anticipated to continue with climate change, population growth, and urbanization. Extreme weather events are the primary cause of increase in frequency and duration of power outages in the U.S. [2]. Annual incurred loss to the U.S. economy by weather-related outages is estimated at \$25 to \$70 billion [3]. The total annual cost of outages to utility customers may increase to over \$480 billion during the 2080-2099 period if aggressive grid resilience enhancement strategies are not implemented [2], [4]. Among the primary systems within the power grid, improving the resilience of distribution networks requires significant attention, as they are highly vulnerable to extreme weather and climate hazards. In fact, they account for 92% of all electric service interruptions in the U.S. [2], [3]. For example, Hurricane Wilma and Hurricane Katrina damaged 12,400 and 72,500 utility poles, respectively [5], [6].

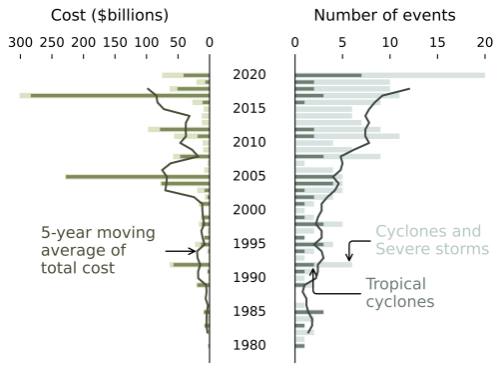

Fig. 1. Temporal pattern of the number of severe storms and tropical cyclones as well as their estimated costs in the U.S. (Data from [1])

Unlike reliability, which is concerned with grid performance under likely events, resilience is a comprehensive metric that measures system performance under shocks. In addition, resilience is concerned with all phases of system response relative to the time of the shock and the associated decisions. Therefore, resilience is a capable metric as the basis for risk management of the power grid against weather and climate extremes. Among key resilience improving strategies is grid hardening measures – a set of proactive actions that improve robustness and resistance of the system to shocks. There are several hardening strategies, such as replacement or reinforcement of utility poles (e.g., [7]–[9]) or vegetation management (e.g., [8]). Hardening distribution lines by replacing deteriorated utility poles is an effective strategy as it mitigates the likelihood of damage and consequently disruptions in the system [8], [10]. In fact, utility pole replacement is recommended by the U.S. National Electric

Safety Code (NESC) [11] and is followed by over 90% of U.S. states [12]. While hardening measures are taken prior to hazard occurrence, operational measures – another category of resilience enhancement strategies – are mainly control-based reactive actions designed to expedite system restoration after disruptions [13], [14]. Among operational measures for the grid repair and restoration phase, developing the capability to reconfigure the topology of the system empowered by controllable distributed generation (DG) units and automatic switches as well as power flow adjustments have gained attention (e.g., [9], [14]–[17]). DGs and switches placed in a distribution system can actively alter the topology after a disastrous event [16], allowing energy consumers in supported areas to continue to receive electricity. Expediting the repair process by routing repair crews is another key operational measure for improving resilience (e.g., [13]–[15], [18]).

Resources for resilience enhancement, however, are limited; therefore, optimization is needed to allocate resources such that the benefits are maximized in the face of constraints. As the resilience maximization problem involves various uncertainties, stochastic optimization (e.g., [15], [19]–[21]) or robust optimization (e.g., [7], [8]) are often used. Although stochastic optimization is a widely accepted approach for optimization under uncertainty, it faces a few practical challenges, including the need for accurate probability distribution for all random variables and high computational complexity, among others [22], [23]. Another limitation relates to the objective of stochastic optimization, which is to increase the expected resilience. However, decision-making based on expected value of performance may leave the system unprepared for extreme events, especially when the problem contains high levels of uncertainty. To address these limitations, a few studies formulated resilience enhancement of distribution systems as a robust optimization problem (e.g., [7], [8]). However, robust optimization models the uncertainty as a deterministic set, and it does not leverage probabilistic information. Furthermore, the solution of robust optimization is based on the worst-case scenario, and therefore the solutions are prone to over-conservatism, which may not be an effective use of limited resources. To overcome these limitations, cutting-edge approaches for optimization under uncertainty have been developed to take advantage of the synergies of the two techniques (e.g., [24], [25]). Advances offered by stochastic robust optimization have increased interests in this group of techniques for decision-making under uncertainty for power systems (e.g., [25], [26]).

This study presents a new resilience decision framework based on a stochastic robust optimization (SRO) model – a synergy of stochastic and robust optimization techniques – for resilience enhancement of power distribution systems facing risks of uncertain hurricane events. The key objective of the proposed resilience planning framework is to determine optimal hardening measures. Although a few studies explored hardening strategies (e.g., [7]–[9]), they largely neglected impacts of operational measures on hardening decisions. However, to obtain optimal hardening solutions for a smart distribution system, impacts of system operation and repair after disruptions on hardening should be properly simulated. For example, in a smart distribution system equipped with tie switches and DG units, critical lines – defined as those whose failure causes the most load shedding in the system – can be different from critical lines in the same but radial system that is not reconfigurable. Therefore, if topology reconfiguration feature is neglected in simulating system operation during and post-event, the optimal hardening solutions may be different compared to the case where reconfiguration is considered. As another example, if a resilience planning framework neglects optimal repair decisions in simulating post-event recovery, it returns a longer duration of outages as a result of disruptions, and consequently overestimates load shedding cost caused by disruptions. Therefore, it is likely that the framework returns non-optimal hardening solutions.

To address the existing gaps, the proposed SRO model determines optimal hardening decisions by modeling key stages of resilience response for an active distribution system that is equipped with controllable DG units and switches. According to the SRO model, hardening decisions are determined considering that the operator makes optimal system operation and repair decisions during and after disruptions in the system. More specifically, the SRO model determines hardening as a proactive pre-event solution, while properly simulating system operation and repair after disruptions by optimizing topology reconfiguration, power flow in the system, and repair crew scheduling as emergency decisions. The proposed SRO model captures the complexities of decision making for actions prior to, during, and after hazards through four stages: pre-event hardening, hurricane shock to the grid, self-healing based on reconfiguration, and recovery process. In the SRO model, hardening decisions are made in advance prior to the event to reduce future possible damages to the system. The shock stage is modeled such that it searches for the expected worst-case scenario of damage to the system given a specified uncertainty budget, while the system operator adjusts power flow to minimize the load shedding cost. In this context, the shock stage is modeled as a bilevel programming, which is transformed to its equivalent single level mixed-integer linear programming (MILP). Once the damages to the system are known as the result of the shock stage, the system enters the self-healing stage to partially restore the load shedding by reconfiguring the topology. In this stage, switches are operated, and power flow is adjusted simultaneously to minimize the load shedding cost. The self-healing stage is modeled as a single level MILP. In the post-hazard period, the interdependency of recovery and system operation decisions are modeled to determine the set of actions that quickly bring back the system to the fully functional state with minimum cost of repairing and traveling between damaged lines, while reducing the load shedding cost by adaptively adjusting the power flow and changing the topology of the system. The recovery process is modeled as a single stage stochastic optimization, which is transformed to its certainty equivalence MILP. A solution strategy based on a differential evolution algorithm and a mixed-integer solver is designed to solve the multi-stage SRO model.

The key contributions of the paper include:

i. We propose a novel multi-stage SRO model as a decision-making tool for smart distribution systems. The model identifies optimal hardening decisions while capturing impacts of operational resilience measures on the decision-making process.

ii. The SRO model provides optimal hardening decisions at a high-resolution for individual poles, while existing frameworks often lump these decisions to a hardening strategy for all poles in each distribution line. Making hardening decisions for a line yields sub-optimal and impractical solutions because it is highly likely that a line includes utility poles with different levels of vulnerability, and it is not effective to do nothing for all poles in a line or replace all of them. The proposed SRO model also includes a new MILP model for the recovery process, which co-optimizes the repair, reconfiguration, and power flow adjustments.

iii. We integrate the probabilistic properties of hurricanes with the uncertain performance of overhead utility structures to develop a new uncertainty set in the proposed SRO model, whereas existing robust optimization-based approaches either neglect probabilistic performance of components or return conservative solutions based on a deterministic pre-defined extreme event. The proposed uncertainty set is established based on the hazard vulnerability of the system and incorporates multi-dimensional fragility models of components. This way, the SRO model releases the common inaccurate assumption that a hardened line does not fail in an extreme event. We also develop a novel probabilistic model for estimating the expected repair time of each damaged line in the system, which is derived based on the Poisson Binomial distribution and Bayes theorem.

## II. MATHEMATICAL FORMULATION

In this section, first, the general form of the proposed multi-stage SRO model is presented. Then, the stages of this model are elaborated in the subsequent subsections.

### A. General optimization model

To effectively improve the resilience of power distribution systems via hardening measures, we present a SRO model where the objective is to minimize the total cost of hazard impacts and response including costs of hardening, incurred damage to the system, and recovery of the system. The SRO model determines optimal hardening policies given a limited budget to defend against the expected worst-case scenario of damage over feasible hurricane events. To simulate the operation of a smart distribution system, optimal corrective actions, including reconfiguration, power flow adjustments, and repair, are implemented to restore the system during and after the hurricane event occurs. In fact, the proposed SRO model innovatively integrates optimal operation of a distribution system into the hardening planning, whereas previous studies often neglect impacts of operational measures on hardening policies. Although impacts of several key operational measures are considered in this study, future research is needed to integrate other operational measures, such as dynamic formation of microgrids. The proposed SRO model is developed based on a few assumptions which can be relaxed in future studies. For example, it is assumed that the hurricane-induced damages to the system occur in a short time interval, and switches in the system are not timely operated within this interval. It is also assumed that the information about out-of-service lines is fully available shortly after the damage caused by hurricane. Another assumption of the SRO model is that the distribution power flow is modeled using the linearized version of DistFlow equations.

The general form of the SRO model is as follows:

$$\min_{x \in \mathbb{X}} \left\{ \mathcal{C}^h + \mathbb{E}_{P(\xi)} \left[ \max_{u \in \mathbb{U}(x,\xi)} \min_{z \in \mathbb{O}(u)} \mathcal{C}^d \right] + \min_{\{z,s\} \in \mathbb{O}(u)} \mathcal{C}^{ld} \right. $$
$$\left. + \mathbb{E}_{P(\xi)} \left[ \min_{r \in \mathbb{R}(x,u,\xi), \{z,s\} \in \mathbb{O}(u)} \mathcal{C}^r \right] \right\} \quad (1)$$

where $x$, $u$, $z$, $s$, and $r$ indicate variables for pole hardening decisions, line damages, power flow, switches, and recovery decisions, respectively. $\mathbb{X}$, $\mathbb{U}$, $\mathbb{O}$, and $\mathbb{R}$ denote the set of feasible hardening decisions, uncertainty set, feasible set of system operation, and feasible set of recovery, respectively. $P(\xi)$ represents the probabilistic information of the weather-related uncertain variables, including wind speed and direction. $\mathcal{C}^h$, $\mathcal{C}^d$, $\mathcal{C}^{ld}$, and $\mathcal{C}^r$ are the cost of hardening, damage, load shedding, and recovery, respectively. The SRO model is developed such that each component in the model describes a specific stage of the resilience curve presented in Fig. 2. The considered chronology of decisions and impacts for grid resilience is as follows. In the first stage, pole hardening decisions are considered as proactive actions that are applied to the system during $t_{pre-event}$. Then, in the second stage, the hardened system faces an extreme event at $t_e$. In this stage, distribution lines are failed such that their failure leads to the maximum level of load shedding. The system operator immediately responds to the disruption by adjusting power flow in the damaged system to minimize the progression of load shedding. This stage (*i.e.*, max-min model) corresponds to the part of the resilience curve where the performance of the system degrades rapidly. It is assumed that the damage caused by hurricane and the corresponding response of the system operator occur in a short time interval after $t_e$. It is also assumed that the response of the system operator in the second stage is limited to adjusting power flow and the topology is not altered by the operator in this period following the procedures in practice [7], [16].

In the third stage, the objective is to minimize the cost of load shedding by simultaneously reconfiguring the system topology via switches and adjusting flow in the system through optimal power flow decisions. It should be noted that deploying a reconfiguration plan requires information about out-of-service lines. To locate faulty lines, the system operator should perform damage assessment [14], [27]. This assessment can be conducted using different techniques, such as fault identification algorithms, dispatching field assessors, and aerial survey [14]. In this study, it is assumed that the information about out-of-service lines is available after the second stage. Thus, as it can be seen in Fig. 2, the third stage starts in a short time interval after the second stage. However, this assumption can be easily updated if an estimate of the duration of damage assessment is available.

The repair and restoration often start after the hurricane passes, and it can be delayed for many reasons, such as safety considerations, assessment of the state of conditions, and coordination with other utilities and government organizations [28]. However, these sources of uncertainty are not considered in this study, and $t_r$ in Fig. 2 indicates the time that hurricane passes, and the recovery process of the system starts. The recovery stage co-optimizes the repair crew scheduling, system topology reconfiguration, and power flow in the system. More specifically, the objective is to minimize the costs of repairing, traveling between damaged lines, and load shedding by determining the interdependent optimal repair crew schedules, power flow and switch variables.

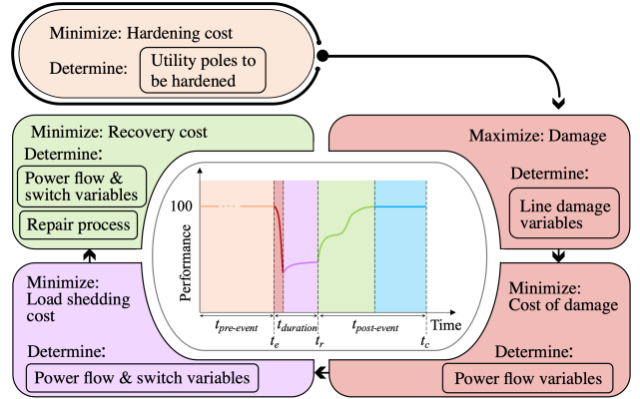

Fig. 2. Flowchart of the proposed resilience enhancement framework

### B. First Stage: Pre-event Hardening

Unlike prior studies where the hardening decision for a line was limited to whether to replace all the poles in the line, the hardening decision for each line in the proposed approach is modeled with an integer variable that specifies the total number of pole replacements in that line. Thus, the cost associated with the hardening is:

$$\mathcal{C}^h(x) = \sum_{(i,j) \in \Omega_L} x_{ij} \, c^h \quad (2)$$

where $x_{ij}$ indicates the total number of replaced poles in line $(i,j)$. $c^h$ denotes the cost coefficient of a utility pole replacement, which can be estimated as the total sum of costs associated with replacing a single wood utility pole with the new one. A greedy strategy is implemented for replacing the poles per line. In this approach, all poles in line $(i,j)$ are ranked based on their failure probability, and $x_{ij}$ poles with the highest probability of failure are replaced with new poles. It is commonly assumed that hardened lines cannot fail in the face of extreme events (*e.g.*, [7], [29]). This assumption, however, is not realistic. As an example, fragility models of class 3 wood poles developed in [30] indicate that the failure probability of a new pole can reach 0.113 for a wind speed of 150 mph. To relax this assumption, replacing aged utility poles with new ones from the same class is formulated to decrease the failure probability of poles, and consequently to reduce the probability of failure of lines. The decrease in pole failure probability is derived using multi-dimensional pole fragility models in [30]. As the budget and resources are limited, constraint (3) is defined to limit the total number of pole replacements. The feasible set (or constraint set) of hardening is as follows:

$$\mathbb{X} = \left\{ x \,\middle|\, \sum_{(i,j) \in \Omega_L} x_{ij} \leq B_H \right. \quad (3)$$

$$\left. x_{ij} \in \{0, 1, \dots, |\psi_{ij}|\}, \quad \forall (i,j) \in \Omega_L \right\} \quad (4)$$

where $x$ denotes pole hardening decision variables and $B_H$ is the hardening budget. Constraint (4) indicates possible hardening decisions, where $\psi_{ij}$ is a set that includes information of all utility poles in line $(i,j)$.

### C. Second Stage: Hurricane Shock to the Grid

The purpose of the shock stage is to model hurricane impacts on a distribution system and the immediate ability of the system to absorb this shock. This stage is commonly modeled as a bilevel robust optimization where the objective of the outer level is to maximize the damage to reach the worst-case scenario of damage, while in the inner level, the system operator implements feasible corrective actions to restore the system after the shock occurs. It is worth noting that natural hazards do not aim at maximizing the damage in reality, and bilevel robust optimization models were initially tailored for man-made attacks. However, in recent years, these models have been widely used in planning for power systems under natural hazards due to their practicality and effectiveness in preparing system operators to plan for unexpected events (*e.g.*, [7], [8], [16], [29]). Although bilevel robust optimization have been shown advantageous, they may lead to overly conservative solutions. In existing bilevel robust models, the outer level seeks to find the worst-case scenario of damage based on a deterministic pre-defined event. Thus, provided solutions highly depend on the assumed event. For example, if the pre-defined event is very rare, resilience enhancement solutions will be extremely conservative. On the other hand, assuming a highly frequent event will lead to unsafe solutions. To address this limitation, this study models the shock stage as a bilevel stochastic robust problem (second term in (1)). The outer level determines the uncertain variables associated with line damage such that $\mathcal{C}^d$ is maximized, which results in the expected worst-case scenario of damage cost. To elaborate more, the outer level contains two sets of uncertain variables, including weather-related uncertain variables (*i.e.*, $\boldsymbol{\xi}$) and line damage uncertain variables (*i.e.*, $\boldsymbol{u}$). In this level, $\boldsymbol{u}$ is determined such that the expected worst-case scenario of damage cost is achieved, where the expectation is taken with respect to $\boldsymbol{\xi}$. The inner problem determines power flow decision variables to adjust the power flow in the damaged system immediately after the disruption, and consequently minimize $\mathcal{C}^d$.

The damage cost $\mathcal{C}^d$ in (1) is expanded in (5) where the first term denotes the immediate load shedding cost in the system due to out-of-service lines. The second term in (5) aims at increasing the number of failed lines in the system when there are no changes in the immediate load shedding and the uncertainty budget is not entirely used. To elaborate more, there can be some cases where an extreme event disconnects all power generation sources from the system. In these cases, the immediate load shedding cost is at its maximum level, therefore, if we do not include the second term in (5), the shock model does not increase the number of failed lines even if the entire uncertainty budget is not used. However, having more failed lines can impact the recovery stage as it results in a longer recovery process, which consequently leads to a higher incurred cost. In other words, the second term in (5) forces the shock model to increase the physical damage to the system (*i.e.*, the number of failed lines in the system) when there are no changes in the immediate load shedding and the uncertainty budget is not entirely used.

$$\mathcal{C}^d(\boldsymbol{\rho^{ld}}, \boldsymbol{u}) = \sum_{t \in \{t_e\}} \sum_{j \in \Omega_B} c_{j,t}^{ld} \rho_{j,t}^{ld} P_{j,t}^{ld} + \sum_{t \in \{t_e\}} \sum_{(i,j) \in \Omega_L} u_{ij,t} \varepsilon_1 \quad (5)$$

$c_{j,t}^{ld}$, $\rho_{j,t}^{ld}$, $P_{j,t}^{ld}$ in (5) indicate cost coefficient of load shedding, ratio of load shedding, and active load demand at bus $j$ and time $t$, respectively. The ratio of load shedding falls within [0,1], where $\rho_{j,t}^{ld} = 1$ if all loads at bus $j$ and time $t$ are shed, and it is zero if there is no load shedding at bus $j$ and time $t$. $\varepsilon_1$ is a significantly small value, and $u_{ij,t}$ denotes the binary line damage variable. $u_{ij,t} = 1$ if the line $(i,j)$ at time $t$ is damaged, and it is zero otherwise. As $\varepsilon_1$ is a significantly small value, the second term in (5) does not increase the cost of damage; however, it forces the shock model to increase the physical damage to the system when there are no changes in the immediate load shedding cost and the uncertainty budget is not entirely used. In this study, it is assumed that the damage caused by hurricane occurs at time $t_e$, and each time step is considered to be one hour. Therefore, the subscript $t$ is omitted in subsections 1, 3, and 4 of Section II.C for ease of exposition.

We transform the bilevel damage cost estimation problem to an equivalent single-level bilinear programming problem, and subsequently linearize the bilinear program using big-M method [8], [31]. To elaborate the details of this transformation, this section first presents the constraints of the outer and inner problems. Then, the dual of the inner problem and the entire transformed problem are introduced.

#### 1) Outer problem

Given an uncertainty set, the outer problem is designed to determine the lines whose failure results in the maximum $\mathcal{C}^d$. In this process, weather-related uncertain variables (*i.e.*, $\boldsymbol{\xi}$) are treated as probabilistic in the sense that their distribution is known, whereas probabilistic information of other uncertain variables (*i.e.*, $\boldsymbol{u}$) is unknown. As $\boldsymbol{\xi}$ affects the bounds and shape of the uncertainty set $\mathbb{U}$, the objective of this level is to reach the expected worst-case cost of damage, where the expectation is taken with respect to the weather-related probabilistic variables. The feasible set of uncertainty (*i.e.*, uncertainty set) is defined as follows:

$$\mathbb{U}(\boldsymbol{x}, \boldsymbol{\xi}) = \left\{ \boldsymbol{u} \,\middle|\, \sum_{(i,j) \in \Omega_L} \left[ -\log_2 P_{ij}^f(x_{ij}, \psi_{ij}, \boldsymbol{\xi}) \right] u_{ij} \leq B_U \right. \quad (6)$$

$$\left. u_{ij} \in \{0,1\}, \quad \forall (i,j) \in \Omega_L \right\} \quad (7)$$

where $\boldsymbol{u}$ is the binary line damage variables. $P_{ij}^f$ and $B_U$ indicate the expected annual failure probability of line $(i,j)$ and uncertainty budget, respectively. It should be noted that $\boldsymbol{\xi}$ includes wind speed $v$ and wind direction $\theta$ as two hurricane hazard characteristics that are significant for the overhead structure performance. Constraint (6) is motivated by the Claude Shannon's information theory [32]. It probabilistically controls the number of failed lines in the system. More specifically, (6) states that the failure of a line with small failure probability is unexpected, therefore, it takes a large portion of the uncertainty budget. On the other hand, failure of a vulnerable line with large failure probability is expected and it takes only a small portion of $B_U$. In this setting, if the failure probability of a line is zero, failure of that line requires an infinite uncertainty budget, and consequently $u_{ij}$ becomes zero if $B_U$ is finite. Commonly, the failure probability of lines is not considered in the definition of uncertainty set (*e.g.*, [7]), therefore, the shock model becomes similar to the traditional $N - K$ worst-case contingency analysis. A few studies related the uncertainty set to the failure probability of lines under a pre-defined extreme wind speed event (*e.g.*, [8], [29]). If the extreme event is treated deterministically, the failure probability of lines, and consequently the wort-case scenario of damage will be returned for that specific pre-defined extreme event. However, in planning problems, the hurricane events should be treated probabilistically, otherwise, the derived decisions based on a single hurricane event highly depend on that event and can be within the range of significantly unsafe to extremely conservative for other events. In this study, the probabilistic information of hurricane hazards is considered to determine the worst-case scenario of damage based on all feasible hurricane events rather than a single event. Therefore, we refer to this worst-case scenario as the expected worst-case scenario throughout the paper as it is derived based on all possible hurricane events. More specifically, the expected failure probability of poles can be estimated by integrating probabilistic hazard and fragility models. Then, the expected failure probability of lines can be calculated using the expected failure probability of pole in order to form the uncertainty set. Consequently, the expected worst-case scenario can be determined given this uncertainty set. Moreover, in existing studies, the same fragility model has been used for the components of overhead lines (*e.g.*, [8], [29]) irrespective of the specific properties of utility poles, such as age, height, span length, and class. Neglecting the effects of these properties may result in inaccurate estimates of failure probability of lines.

As utility poles in a line are connected in series, assuming that the poles fail independently, the expected annual failure probability of line $(i, j)$ is calculated as:

$$P_{ij}^f = 1 - \prod_{m \in \psi_{ij}} \left(1 - p_{m,ij}^f\right) \tag{8}$$

where $p_{m,ij}^f$ is the expected annual failure probability of pole $m$ in line $(i, j)$, and is computed as:

$$p_{m,ij}^f = \iint Pr(F | \Gamma, H, A_C, v, \theta). f_V(v). f_\theta(\theta). dv \, d\theta \tag{9}$$

where $\Gamma$ and $H$ denote the age and height of the pole. $A_C$ is the conductor area, and $v$ and $\theta$ are the wind speed and wind direction, respectively. $Pr(F | \Gamma, H, A_C, v, \theta)$ is the failure probability of the pole given its characteristics together with wind speed and direction. This probability of failure is obtained using the multi-dimensional fragility model in [30]. Moreover, $f_V$ and $f_\theta$ are the probability density functions of wind speed and wind direction on a yearly basis, respectively. Integrating the fragility model of a pole with the joint distribution of annual wind speed and direction returns the annual failure probability of that pole. It is worth noting that age plays an important role in the annual probability of failure, thus, if a pole is replaced with a new one as the result of hardening decisions, the annual failure probability of that pole will reduce significantly.

*2) Inner problem*

In the inner level, the system operator responds to the disruption by adjusting power flow in the damaged distribution system to minimize $C^d$. Linearized DistFlow equations [33] are used to determine the complex power flow and voltage profile in the distribution system. These equations have been extensively implemented to model power flow through radial distribution systems (*e.g.*, [7], [8], [34]). When the system is reconfigured dynamically via switches, the radial configuration is maintained. To enforce radiality at all times, the radiality constraints based on the spanning tree approach in [35] are adopted. It is worth reminding that the second stage of the proposed model does not consider reconfiguration. However, the radiality constraints are included in the feasible set of system operation in this section as this set is the basis of models in the third and fourth stages. Thus, the presented feasible set of system operation in (10)-(27) contains conservation of flow constraints, capacity constraints, and radiality constraints:

$$\mathbb{O}(u) = \left\{ z, s \; \middle| \; \sum_{k \in \phi(j)} P_{kj,t} \right. $$
$$= \sum_{k \in \varphi(j)} P_{jk,t} - P_{j,t}^g + (1 - \rho_{j,t}^{ld}) P_{j,t}^{ld}, \tag{10}$$
$$\forall j \in \Omega_B, t \in \Omega_T$$

$$\sum_{k \in \phi(j)} Q_{kj,t} = \sum_{k \in \varphi(j)} Q_{jk,t} - Q_{j,t}^g + (1 - \rho_{j,t}^{ld}) Q_{j,t}^{ld}, \tag{11}$$
$$\forall j \in \Omega_B, t \in \Omega_T$$

$$0 \le P_{ij,t} \le (1 - y_{ij,t}) P_{ij}^{max}, \qquad \forall (i,j) \in \Omega_L, t \in \Omega_T \tag{12}$$

$$0 \le Q_{ij,t} \le (1 - y_{ij,t}) Q_{ij}^{max}, \qquad \forall (i,j) \in \Omega_L, t \in \Omega_T \tag{13}$$

$$|V_{j,t}| \le |V_{i,t}| - \frac{R_{ij} P_{ij,t} + X_{ij} Q_{ij,t}}{V_0} + y_{ij,t} M_1, \tag{14}$$
$$\forall j \in \Omega_B, i = \phi(j), t \in \Omega_T$$

$$|V_{j,t}| \ge |V_{i,t}| - \frac{R_{ij} P_{ij,t} + X_{ij} Q_{ij,t}}{V_0} - y_{ij,t} M_1, \tag{15}$$
$$\forall j \in \Omega_B, i = \phi(j), t \in \Omega_T$$

$$0 \le P_{j,t}^g \le P_j^{g,max}, \qquad \forall j \in \Omega_B, t \in \Omega_T \tag{16}$$

$$0 \le Q_{j,t}^g \le Q_j^{g,max}, \qquad \forall j \in \Omega_B, t \in \Omega_T \tag{17}$$

$$|V_{j,t}| \ge |V_j|^{min}, \qquad \forall j \in \Omega_B, t \in \Omega_T \tag{18}$$

$$|V_{j,t}| \le |V_j|^{max}, \qquad \forall j \in \Omega_B, t \in \Omega_T \tag{19}$$

$$\rho_{j,t}^{ld} \in [0,1], \qquad \forall j \in \Omega_B, t \in \Omega_T \tag{20}$$

$$y_{ij,t} \le u_{ij,t} + s_{ij,t} - u_{ij,t} s_{ij,t}, \tag{21}$$
$$\forall (i,j) \in \Omega_L, t \in \Omega_T$$

$$y_{ij,t} \ge u_{ij,t} + s_{ij,t} - u_{ij,t} s_{ij,t}, \tag{22}$$
$$\forall (i,j) \in \Omega_L, t \in \Omega_T$$

$$s_{ij,t} \le 0, \qquad \forall (i,j) \notin \Omega_{SW}, t \in \Omega_T \tag{23}$$

$$\beta_{ij,t} + \beta_{ji,t} \le 1 - y_{ij,t}, \qquad \forall (i,j) \in \Omega_L, t \in \Omega_T \tag{24}$$

$$\beta_{ij,t} + \beta_{ji,t} \ge 1 - y_{ij,t}, \qquad \forall (i,j) \in \Omega_L, t \in \Omega_T \tag{25}$$

$$\beta_{ij,t} \le 0, \qquad \forall j \in \Omega_{RB}, i = \phi(j), t \in \Omega_T \tag{26}$$

$$\left. \sum_{k \in \{\phi(j) \cup \varphi(j)\}} \beta_{kj,t} \le 1, \qquad \forall j \in \Omega_B, t \in \Omega_T \right\} \tag{27}$$

where $z$ and $s$ indicate power flow and switch decision variables, respectively. The equality constraints in (10) and (11)

represent conservation of the active and reactive power flow at each bus. Constraints (12) and (13) impose limits on the active and reactive line power flows. These constrains force the power flow of line to be zero if the status of that line becomes disconnected as the result of failure under extreme events and/or opened switch. (14) and (15) represent the voltage level at each bus. Constraints (16) and (17) bound active and reactive power generation of DGs. It is worth noting that DGs can be modeled in four types based on their active and reactive power delivering capacity. A DG unit, such as photovoltaic and fuel cells, can only inject active power. On the other hand, DG units equipped with synchronous compensator can be considered as DGs capable of injective reactive power only. DG units that are based on synchronous machine are capable of injecting both active and reactive power. Induction generators that are used in wind farms can be considered as DGs capable of injecting active power, while consuming reactive power [36], [37]. As it can be seen in constraints (16) and (17), in this study, it is assumed that DGs can inject both active and reactive power. (18) and (19) indicate the voltage bounds. Constraint (20) sets the limits on load shedding ratios. (21) and (22) represent the relation between binary status variables and binary line damage and switch variables. According to these two constraints, the binary status variable of line $(i, j)$ at time $t$, $y_{ij,t}$, becomes 0 when neither the line is damaged nor the switch at the line is open at time $t$. Constraint (23) forces the switch variable becomes 0 for the lines without switch devices. (24)-(27) represent the radiality constraints. Two binary variables $\beta_{ij,t}$ and $\beta_{ji,t}$ are defined to model the spanning tree. $\beta_{ij,t} = 1$ indicates that bus $i$ is the parent bus to child bus $j$ at time $t$. With similar analogy, $\beta_{ji,t} = 1$ represents that bus $j$ is the parent bus to child bus $i$ at time $t$. (24) and (25) indicate that either $\beta_{ij,t}$ or $\beta_{ji,t}$ must be 1 if the line $(i, j)$ is connected at time $t$. Constraint (26) represents that root buses do not have a parent bus. In a radial system, each bus cannot have more than one parent bus. To impose this feature of radial systems, constraint (27) is defined.

Constraints (21) and (22) contain a nonlinear term which is the product of two binary variables. This term is linearized by replacing $u_{ij,t} s_{ij,t}$ with a binary variable $\gamma_{ij,t}$ and adding the following three constraints:

$$\gamma_{ij,t} \leq u_{ij,t} \tag{28}$$

$$\gamma_{ij,t} \leq s_{ij,t} \tag{29}$$

$$\gamma_{ij,t} \geq u_{ij,t} + s_{ij,t} - 1 \tag{30}$$

### 3) Dual of the inner problem

Under the strong duality, primal solutions can be characterized from dual solutions. Thus, by taking the dual of the inner problem, the bilevel shock model is converted to a single maximization problem. As noted earlier, it is assumed that the switches may not be timely operated in the inner level of the shock model. Consequently, the topology will not be changed by system operators in this level. Therefore, as the system is radially operated prior to the extreme event occurrence, the system maintains the radial topology in the shock stage, which indicates that the radiality constraints (24)-(27) are satisfied herein. Moreover, as the switches are not operated in this stage, the binary status variable depends only on the binary line damage variable. Considering these, the feasible set of system operation of the inner problem is simplified to constraints (10)-(20) where $y_{ij,t}$ is replaced with $u_{ij,t}$. It is also worth reminding that the shock model simulates hurricane impacts on a distribution system and the immediate ability of the system to absorb this shock during a single time step (*i.e.*, one hour in this study). Thus, the subscript $t$ is omitted in the rest of this subsection for ease of exposition. To form the corresponding dual problem of the inner problem, its Lagrangian function is defined as follows:

$$
\begin{aligned}
\mathcal{L}(\boldsymbol{z}, \boldsymbol{u}, \boldsymbol{\lambda}, \boldsymbol{\mu}) =& \sum_{j \in \Omega_B} c_j^{ld} \rho_j^{ld} P_j^{ld} + \sum_{(i,j) \in \Omega_L} u_{ij} \varepsilon_1 \\
&+ \sum_{j \in \Omega_B} \mu_1^j \left[ \sum_{k \in \varphi(j)} P_{jk} - P_j^g \right] \\
&+ (1 - \rho_j^{ld}) P_j^{ld} - \sum_{k \in \phi(j)} P_{kj} \\
&+ \sum_{j \in \Omega_B} \mu_2^j \left[ \sum_{k \in \varphi(j)} Q_{jk,t} - Q_{j,t}^g \right] \\
&+ (1 - \rho_{j,t}^{ld}) Q_{j,t}^{ld} - \sum_{k \in \phi(j)} Q_{kj} \\
&+ \sum_{(i,j) \in \Omega_L} \lambda_1^{ij} [ P_{ij} - (1 - u_{ij}) P_{ij}^{max} ] \\
&+ \sum_{(i,j) \in \Omega_L} \lambda_2^{ij} [ Q_{ij} - (1 - u_{ij}) Q_{ij}^{max} ] \\
&+ \sum_{j \in \Omega_B} \lambda_3^j [ |V_j| - |V_i| \\
&+ \frac{R_{ij} P_{ij} + X_{ij} Q_{ij}}{V_0} - u_{ij} M_1 ] \\
&- \sum_{j \in \Omega_B} \lambda_4^j [ |V_j| - |V_i| \\
&+ \frac{R_{ij} P_{ij} + X_{ij} Q_{ij}}{V_0} + u_{ij} M_1 ] \\
&+ \sum_{j \in \Omega_B} \lambda_5^j [ P_j^g - P_j^{g,max} ] \\
&+ \sum_{j \in \Omega_B} \lambda_6^j [ Q_j^g - Q_j^{g,max} ] \\
&- \sum_{j \in \Omega_B} \lambda_7^j [ |V_j| - |V_j|^{min} ] \\
&+ \sum_{j \in \Omega_B} \lambda_8^j [ |V_j| - |V_j|^{max} ] \\
&+ \sum_{j \in \Omega_B} \lambda_9^j [ \rho_j^{ld} - 1 ]
\end{aligned}
\tag{31}
$$

where $\mu_1$ and $\mu_2$ indicate the dual variables of equality constraints (10) and (11), respectively, which are not restricted in sign. $\lambda_1$ to $\lambda_9$ are non-negative dual variables of constraints (12) to (20), respectively. By taking a pointwise minimum of the Lagrangian (*i.e.*, $\mathcal{G}(\boldsymbol{u}, \boldsymbol{\lambda}, \boldsymbol{\mu}) = \min_{\boldsymbol{z}} \mathcal{L}(\boldsymbol{z}, \boldsymbol{u}, \boldsymbol{\lambda}, \boldsymbol{\mu})$), the Lagrange dual function can be obtained as:

$$\mathcal{G}(\boldsymbol{u}, \boldsymbol{\lambda}, \boldsymbol{\mu}) = \sum_{(i,j) \in \Omega_L} u_{ij} \varepsilon_1 + \sum_{j \in \Omega_B} P_j^{ld} \mu_1^j + \sum_{j \in \Omega_B} Q_j^{ld} \mu_2^j$$
$$- \sum_{(i,j) \in \Omega_L} \lambda_1^{ij} (1 - u_{ij}) P_{ij}^{max}$$
$$- \sum_{(i,j) \in \Omega_L} \lambda_2^{ij} (1 - u_{ij}) Q_{ij}^{max}$$
$$- \sum_{j \in \Omega_B} \lambda_3^j u_{ij} M_1 - \sum_{j \in \Omega_B} \lambda_4^j u_{ij} M_1 \qquad (32)$$
$$- \sum_{j \in \Omega_B} \lambda_5^j P_j^{g,max} - \sum_{j \in \Omega_B} \lambda_6^j Q_j^{g,max}$$
$$+ \sum_{j \in \Omega_B} \lambda_7^j |V_j|^{min} - \sum_{j \in \Omega_B} \lambda_8^j |V_j|^{max}$$
$$- \sum_{j \in \Omega_B} \lambda_9^j$$

The Lagrange dual function in (32) contains four bilinear terms which are linearized using big-M method. More specifically, $\lambda_1^{ij}(1 - u_{ij})$, $\lambda_2^{ij}(1 - u_{ij})$, $\lambda_3^j u_{ij} M_1$, and $\lambda_4^j u_{ij} M_1$ are replaced with non-negative variables $\eta_1^{ij}$, $\eta_2^{ij}$, $\eta_3^j$, and $\eta_4^j$, respectively. These transformations add the following constraints to the model:

$$\eta_1^{ij} - \lambda_1^{ij} + M_2 u_{ij} \geq 0, \qquad \forall (i,j) \in \Omega_L \qquad (33)$$

$$\eta_2^{ij} - \lambda_2^{ij} + M_2 u_{ij} \geq 0, \qquad \forall (i,j) \in \Omega_L \qquad (34)$$

$$\eta_3^j - \lambda_3^j M_1 + M_2 (1 - u_{ij}) \geq 0, \ \ \forall j \in \Omega_B, i = \phi(j) \qquad (35)$$

$$\eta_4^j - \lambda_4^j M_1 + M_2 (1 - u_{ij}) \geq 0, \ \ \forall j \in \Omega_B, i = \phi(j) \qquad (36)$$

### 4) Transformed shock model

Under the strong duality, the Karush-Kuhn-Tucker (KKT) conditions are necessary for optimality. Using the results of KKT optimality conditions, the transformed MILP shock model can be summarized as:

$$\max \mathcal{G}(\boldsymbol{u}, \boldsymbol{\lambda}, \boldsymbol{\mu}, \boldsymbol{\eta}) \qquad (37)$$

$$\mu_1^i - \mu_1^j + \lambda_1^{ij} + \frac{R_{ij}}{V_0} \lambda_3^j - \frac{R_{ij}}{V_0} \lambda_4^j = 0,$$
$$\forall j \in \Omega_B, i = \phi(j) \qquad (38)$$

$$\mu_2^i - \mu_2^j + \lambda_2^{ij} + \frac{X_{ij}}{V_0} \lambda_3^j - \frac{X_{ij}}{V_0} \lambda_4^j = 0,$$
$$\forall j \in \Omega_B, i = \phi(j) \qquad (39)$$

$$\lambda_3^j - \lambda_4^j - \sum_{k \in \sigma(j)} \lambda_3^k + \sum_{k \in \varphi(j)} \lambda_4^k - \lambda_7^j + \lambda_8^j = 0,$$
$$\forall j \in \Omega_B \qquad (40)$$

$$-\mu_1^j + \lambda_5^j \geq 0, \quad \forall j \in \Omega_B \qquad (41)$$

$$-\mu_2^j + \lambda_6^j \geq 0, \quad \forall j \in \Omega_B \qquad (42)$$

$$c_j^{ld} P_j^{ld} - \mu_1^j P_j^{ld} - \mu_2^j Q_j^{ld} + \lambda_9^j \geq 0, \qquad \forall j \in \Omega_B \qquad (43)$$

$$\boldsymbol{u} \in \mathbb{U} \qquad (44)$$

$$\boldsymbol{\lambda} \geq \boldsymbol{0}, \boldsymbol{\eta} \geq \boldsymbol{0} \qquad (45)$$

Constraints (33)-(36) $\qquad (46)$

### D. Third Stage: Self-Healing Based on Reconfiguration

Self-healing describes the ability of a distribution system in autonomous service restoration after a disruption [38], [39]. Self-healing based on system reconfiguration is considered as an important feature of a smart grid [16], [39]. This feature has been widely used in active distribution systems for service restoration (e.g.,[14]–[16], [40], [41]). In this study, the self-healing based on reconfiguration starts after the second stage of the SRO model. The purpose of the third stage is to reduce the load shedding in the system once the damage to the system is detected. This stage is a critical step in restoring the power in the system prior to the recovery process because in reality, the initiation of repair can be a time-consuming process and system reconfiguration can restore power loss partially until the recovery process starts. The objective function of this stage is the cost of load shedding in a time interval between the end of the second and beginning of the fourth stages of the SRO model as follows:

$$\mathcal{C}^{ld}(\boldsymbol{\rho^{ld}}) = \sum_{t \in (t_e+1, t_r)} \sum_{j \in \Omega_B} c_{j,t}^{ld} \rho_{j,t}^{ld} P_{j,t}^{ld} \qquad (47)$$

where $t_e$ and $t_r$ are the start time of the event and recovery, respectively. In the third stage, power flow and switch variables are determined simultaneously given the constraints (10)-(30) to optimally adjust the power flow in the damaged system and reconfigure the system.

### E. Fourth Stage: Recovery Process

The recovery stage seeks to model the ability of a distribution system to quickly recover from a disruptive event. Many prior studies (e.g., [8], [14]) consider the repair time in the recovery process deterministically. A few studies (e.g., [8]) assume that repairing a line starts instantaneously after its failure. This assumption implies the availability of unlimited resources and repair crew. In addition, this assumption does not represent the reality where the repair and restoration often start after the hurricane passes [28]. To address these limitations, the recovery process is modeled as a single stage stochastic optimization (the fourth term in (1)) and further transformed to its equivalence MILP. The objective is to minimize $\mathcal{C}^r$ by co-optimizing repair crew scheduling and system operation. More specifically, the system operator simultaneously reacts to the repaired lines by adjusting power flow and reconfiguring the system topology. $\mathcal{C}^r$ includes the load shedding cost due to out-of-service lines over the recovery time, cost of repairing the damaged lines, and the cost of traveling between damaged lines as follows:

$$\mathcal{C}^r(\boldsymbol{\rho^{ld}}, \boldsymbol{r}, \boldsymbol{\delta}) = \sum_{j \in \Omega_B} \sum_{t \in \Omega_T} c_{j,t}^{ld} \rho_{j,t}^{ld} P_{j,t}^L$$
$$+ \sum_{(i,j) \in \Omega_F} \sum_{t \in \Omega_T} c_{ij,t}^{rp} r_{ij,t} \qquad (48)$$
$$+ \sum_{(i,j) \in \Omega_F} \sum_{t \in \{\Omega_T \setminus t_r\}} c^{tr} \delta_{ij,t}$$

where $c_{j,t}^{ld}$, $c_{ij,t}^{rp}$, and $c^{tr}$ are the cost coefficients corresponding to load shedding at bus $j$ and time $t$, repairing line $(i,j)$ at time $t$ with a single repair crew, and traveling between two damaged lines, respectively. The set of recovery time is considered as $\Omega_T = \{t_r, ..., t_c\}$, where $t_r$ and $t_c$ are the start time of the recovery, and control time, respectively. As the feasible set of system operation is explained in Section II.C.2, the remainder of this section only introduces the constraints and variables of the repair and traveling during the recovery process. The feasible set of repair actions is proposed as:

$$\mathbb{R}(\boldsymbol{x}, \boldsymbol{u}, \boldsymbol{\xi}) = \left\{ \boldsymbol{r} \,\middle|\, \sum_{(i,j) \in \Omega_F} r_{ij,t} \leq n, \qquad \forall\, t \in \Omega_T \right. \tag{49}$$

$$\frac{1}{r_{ij}(x_{ij}, \psi_{ij}, \boldsymbol{\xi})} \sum_{m=1}^{t} r_{ij,m} \geq 1 - u_{ij,t}, \\ \forall (i,j) \in \Omega_F, t \in \Omega_T \tag{50}$$

$$\sum_{t \in \Omega_T} r_{ij,t} \geq r_{ij}(x_{ij}, \psi_{ij}, \boldsymbol{\xi}), \qquad \forall (i,j) \in \Omega_F \tag{51}$$

$$r_{ij,t-1} - r_{ij,t} \leq M_3(1 - \delta_{ij,t}) - \varepsilon_2, \\ \forall (i,j) \in \Omega_F, t \in \{\Omega_T \backslash t_r\} \tag{52}$$

$$r_{ij,t-1} - r_{ij,t} \geq -M_3 \delta_{ij,t}, \\ \forall (i,j) \in \Omega_F, t \in \{\Omega_T \backslash t_r\} \tag{53}$$

$$\delta_{ij,t} \in \{0,1\}, \quad \forall (i,j) \in \Omega_F, t \in \Omega_T \tag{54}$$

$$\left. r_{ij,t} \in \{0,1,\dots,n\}, \quad \forall (i,j) \in \Omega_F, t \in \Omega_T \right\} \tag{55}$$

where $\boldsymbol{r}$ denotes the recovery decision variables. More specifically, $r_{ij,t}$ is the number of repair crews repairing line $(i,j)$ at time $t$. $n$ represents the total number of repair crews. $r_{ij}$ is the uncertain repair time of line $(i,j)$ in crew-hours. In other words, $r_{ij}$ indicates the time (in hours) that a single crew should spend to repair the failed line $(i,j)$, where the set of failed lines $\Omega_F$, is specified as the result of the shock model. $M_3$ and $\varepsilon_2$ are big and small numbers, respectively. $\boldsymbol{\delta}$ denotes the travel variables, which depends on $\boldsymbol{r}$. Constraint (49) indicates that the number of repair crews repairing all lines at time $t$ cannot exceed the total number of repair crews. Constraint (50) forces the binary line damage variable of the failed line $(i,j)$ to remain 1 until the ratio $\sum_{m=1}^{t} r_{ij,m} / r_{ij}$ reaches or exceeds 1. This ratio reaches 1 when the damaged line $(i,j)$ is fully repaired. In other words, constraint (50) represents that the binary line damage variable of the failed line $(i,j)$ can change to 0 once this line is repaired. Constraint (51) indicates that all failed lines should be repaired between the start time of the recovery and the control time. (52) and (53) are defined to determine binary travel variables. More specifically, these two constraints indicate that if $r_{ij,t}$ is greater than $r_{ij,t-1}$, $\delta_{ij,t}$ becomes 1. When $r_{ij,t}$ is greater than $r_{ij,t-1}$, it means that the number of repair crews repairing line $(i,j)$ at time $t$ is increased compared to the time $t-1$. To increase the number of repair crew at a failed line, repair crews should travel to that line. The binary travel variables are determined following this process. (54) and (55) set limits on travel and recovery decision variables, respectively.

In the proposed model, the uncertainty in $r_{ij}$ originates from the fact that the repair time of a failed line depends on the degree of damage of that line. In fact, as the poles in a line are connected in series, even failure of one pole leads to the failure of that line. Thus, a failed line may contain different number of damaged poles, ranging from one to all the poles in that line. Moreover, the number of poles in a line may vary from other lines. Therefore, assuming the same repair time for all lines in a system as done in most of the prior resilience studies is not accurate. On the other hand, assuming that the repair time for each line is proportional to its length may overestimate the repair time. To properly estimate the required time for repairing a failed line, this study develops a probabilistic repair time model. This model relates the failure probability of poles to the uncertain repair time $r_{ij}$ by assuming that pole failures in each line are independent. Considering the independency of failure of poles in line $(i,j)$, failure of each pole follows a Bernoulli distribution with probability $p_{m,ij}^f$ in (9). Therefore, the total number of pole failures per line follows a Binomial distribution. However, as the failure probability of poles in each line may be different, the Bernoulli trials are not identically distributed. In fact, the total number of pole failures follows the Poisson Binomial distribution that is the discrete probability distribution of a sum of independent Bernoulli trials that are not necessarily identically distributed:

$$N_{ij}^f \sim PB\left( \sum_{m \in \psi_{ij}} p_{m,ij}^f, \sum_{m \in \psi_{ij}} p_{m,ij}^f (1 - p_{m,ij}^f) \right) \tag{56}$$

where $N_{ij}^f$ represents total number of pole failures in line $(i,j)$. Thus, the repair time of line $(i,j)$ in crew-hours (i.e., $r_{ij}$) can be estimated as the product of $N_{ij}^f$ and the required time for repairing a single pole by a repair crew. However, it should be noted that $N_{ij}^f$ in (56) presents the total number of pole failures in line $(i,j)$ without knowing whether the line $(i,j)$ is failed or not. However, when the recovery process starts, the failed lines are revealed. Considering that for a failed line at least one of its poles is failed (i.e., $N_{ij}^f \geq 1$), the distribution of $N_{ij}^f$ in (56) for a failed line is updated using the Bayes theorem. Based on the updated distribution, the probabilistic properties of $r_{ij}$ for those lines can be derived.

One approach to solve this recovery model is to draw many random realizations for $r_{ij}$ using sampling, however, this approach is time-consuming. An alternative method is to substitute the random variables with their expected values, and transform the stochastic optimization to its certainty equivalence problem [42]. Following this approximation, herein, the uncertain variable $r_{ij}$ is replaced with its expected value, and the recovery model is transformed to its certainty equivalence problem. The expected repair time of the failed line $(i,j)$ in crew-hours is calculated as follows:

$$\mathbb{E}[r_{ij}] = \frac{h^r \times \sum_{m \in \psi_{ij}} p_{m,ij}^f}{1 - \prod_{m \in \psi_{ij}} (1 - p_{m,ij}^f)} \tag{57}$$

where $h^r$ denotes the total number of hours that is needed for a single crew to repair a pole.

## III. SOLUTION STRATEGY

The hurricane shock (i.e., second stage), self-healing based on reconfiguration (i.e., third stage), and recovery (i.e., fourth stage) models are MILPs, which are implemented in Python and solved using Gurobi solver. In the proposed multi-stage SRO model, some of the decision variables in the first stage (i.e., hardening model) and the second and fourth stages are coupled, i.e., $\boldsymbol{x}$, $\boldsymbol{u}$, and $\boldsymbol{r}$. More specifically, the uncertainty set of damages $\mathbb{U}$ and feasible set of repair decisions $\mathbb{R}$ depend on the selected hardening strategy. Consequently, binary line damage and discrete recovery decision variables are coupled with the selected hardening strategy. Therefore, the first stage and the other stages cannot be directly decomposed to a master problem and sub-problems, and consequently conventional solution algorithms, such as traditional Benders decomposition and column-and-constraint generation, are no longer applicable [8], [29], [43]. To tackle this challenge, the pre-event hardening stage is implemented in Python and solved using the D-ICDE algorithm that is designed for constrained optimization problem with discrete variables [44]. D-ICDE is a differential evolutionary (DE) algorithm, which is an extension of the improved $(\mu + \lambda)$-constrained differential evolution (ICDE) proposed by Jia et al. [45]. Similar to other metaheuristics, D-ICDE trades off local search and global exploration, therefore, there is no guarantee that the solution is the global optimum [46]. The idea behind using metaheuristics is to have an algorithm that is able to find a good-quality solution in an acceptable timescale [46]. Among metaheuristics, DE algorithms are capable of returning optimum or near optimum solutions, while they are simple to implement and ideally designed for parallel computing [47]. Considering these, D-ICDE is used in this study as it is designed for discrete variables, which is the case for the proposed hardening model, and it has the features of the ICDE to solve constrained optimization problems [21], [44], [45].

Similar to other differential evolution algorithms, D-ICDE consists of four main phases, including initialization, mutation, crossover, and selection. D-ICDE has three main hyperparameters, including scale factor ($F_s$), crossover rate ($C_r$), and size of population ($n_p$). With these three main hyperparameters properly configured, an initial parent population of $n_p$ individuals ($P_0^p$) is generated by randomly sampling from the hardening search space. At each iteration, a new generation replaces the old population through the processes of mutation, crossover, and selection. More specifically, at each iteration, for each individual in the parent population, three offspring are generated following three different mutation and crossover strategies. More details on these strategies can be found in [44], [45]. After the mutation and crossover phases, an offspring population of $3 \times n_p$ individuals ($Q_g^p$) is generated. Then, the parent and offspring populations are combined to create a combined population ($H_g^p$). Subsequently, the objective function value and the degree of constraint violation for all individuals in $H_g^p$ are evaluated. Herein, the objective function value refers to the summation of $C^h$ and the optimized cost function value of the shock and recovery models. The degree of constraint violation can also be computed based on constraint (3). Finally, $n_p$ individuals are selected using an archiving-based adaptive tradeoff model (ArATM) [45] to constitute the next parent population ($P_{g+1}^p$). ArATM is implemented in this study as the selection phase because it can appropriately make a tradeoff between the objective function value and the degree of constraint violation. More information on ArATM can be found in [21], [44], [45]. This iterative process continues for $n_g$ generations to identify the optimal hardening strategy. The implemented solution strategy is outlined in Algorithm 1.

---

**Algorithm 1** Solution Strategy

1: Set the generation number, $g$, to zero ($g = 0$)
2: Initialize main hyperparameters, $F_s$, $C_r$, and $n_p$
3: Initialize parent population by randomly sampling $n_p$ individuals from the hardening search space: $P_0^p = \{x_1, x_2, ..., x_{n_p}\}$
4: **for** $g = 0$: $n_g$
5:     Generate $Q_g^p$ following the mutation and crossover strategies
6:     Constitute $H_g^p$ with size $4n_p (H_g^p = P_g^p \cup Q_g^p)$
7:     **for** $i = 1$: $4n_p$
8:         Set the $x_i$ as the hardening strategy
9:         Solve shock model using (37)-(46) to find $u$, $z$
10:         Solve self-healing model by (10)-(30), (47) to find $z$, $s$
11:         Solve recovery model by (10)-(30), (48)-(55) to find $r$, $\delta$, $z$, $s$
12:         Evaluate objective function value for $x_i$ by (2), (5), (47), (48)
13:         Compute degree of constraint violation for $x_i$ using (3)
14:     **end for**
15:     Select $n_p$ individuals from $H_g^p$ by ArATM to form $P_{g+1}^p$
16: **end for**
17: Return the optimal policy $x^*$ as the hardening strategy

---

## IV. CASE STUDY

The proposed model and the implemented solution algorithm are applied to the IEEE 33-bus distribution system with four DG units and the 118-bus system introduced by Zhang et al. [48] with ten DGs. The systems are assumed to be located in Harris County, TX, U.S. Information about the systems and their components is provided next.

### A. System information

The 33-bus system includes one DG at the root bus with the active and reactive power capacity of 10-MW and 10-MVar, respectively, and three DGs with the active and reactive power capacity of 2-MW and 2-MVar at nodes 7, 12, and 27. This system has five lines that are equipped with switches (see Fig. 4). The second case study is the 118-bus system which contains one DG at the root bus with the active and reactive power capacity of 10-MW and 10-MVar, and nine DG units with the active and reactive power capacity of 2-MW and 2-MVar, respectively, at nodes 2, 7, 15, 40, 62, 74, 82, 94, and 108. This case study has fifteen lines that are equipped with switches (see Fig. 7). Information about the utility poles in the studied systems is not available. Therefore, we randomly select important features of utility poles in these systems. Random realizations of class and height of poles are generated based on a database of a real-world distribution system located in southeast of U.S. [30]. The distribution of class and corresponding height of poles are presented in Table I. The age of the poles is assumed to follow a lognormal distribution [20], [49] with a mean and standard deviation of 45 and 20 years, respectively. In the studied distribution systems, span length follows a lognormal distribution with the mean of 43.9 meters and coefficient of variation of 0.25 [50]. The total number of poles per line are determined based on the length of the line and random realizations of span length. Herein, the studied 33-bus and 118-bus systems consist of 485 and 1841 wood utility poles, respectively.

As shown in (9), the annual failure probability of poles is a function of the class, age, and height of poles, conductor area, and probabilistic information of wind speed and direction. The conductor area is calculated as the product of the conductor diameter, span length, and the number of conductors carried by poles. Here, it is assumed that all poles carry three conductors with the diameter of 0.563 in. Probabilistic information of wind speed and wind direction required to obtain the probability of failure are elaborated below.

TABLE I. DISTRIBUTION OF CLASS AND HEIGHT OF POLES

| Class | Probability mass function | Height (m) | |
|---|---|---|---|
| | | Lognormal distribution | |
| | | Mean | Standard deviation |
| 1 | 0.0075 | 21.6 | 7.1 |
| 2 | 0.0527 | 16.1 | 2.9 |
| 3 | 0.2043 | 13.6 | 1.4 |
| 4 | 0.1925 | 12.3 | 0.7 |
| 5 | 0.5137 | 10.9 | 1.3 |
| 6 | 0.0176 | 9.5 | 0.9 |
| 7 | 0.0117 | 9.2 | 0.7 |

### B. Hurricane hazard

As hurricanes are region-specific, probabilistic hurricane hazard models should be developed specific to the region of interest. These models probabilistically describe hazard characteristics that are significant for the infrastructure performance, *e.g.*, wind speed and direction. Peterka and Shahid [51] showed that a Weibull distribution provides a reasonable fit to the hurricane wind speed in the southern U.S. Darestani et al. [52] found that the annual wind speed (in mph) in Harris County follows a Weibull distribution with scale parameter of 45.4 and shape parameter of 1.2. Moreover, it is assumed that wind direction is uniformly distributed [52].

### C. Cost coefficients

Four cost coefficients are needed for the proposed SRO model, including cost of pole replacement, load shedding, repairing a damaged line, and traveling between damaged lines. These coefficients are presented in Table II [53]–[55].

TABLE II. COST COEFFICIENTS USED IN THIS STUDY

| Description | Notation | Cost |
|---|---|---|
| Replacing a utility pole | $c^h$ | $3350 /pole |
| Load shedding | $c^{ld}$ | $17.4 kW/hr |
| Repairing a damaged line | $c^{rp}$ | $560 crew/hr |
| Traveling between two lines | $c^{tr}$ | $10 |

### D. Results for IEEE 33-bus system

The SRO model is applied to the introduced 33-bus system given a hardening budget (*i.e.*, $B_H$) of 50 poles and uncertainty budget (*i.e.*, $B_U$) of 10. The required time for a repair crew to repair a failed pole (*i.e.*, $h^r$) is considered as 9 hours [55]. The total number of repair crews (*i.e.*, $n$) is set to 3. The duration of hurricane (*i.e.*, $t_r - t_e$) and control time (*i.e.*, $t_c$) are considered as 24 and 72 hours, respectively. In the solution stage, the number of individuals (*i.e.*, $n_p$), scale factor (*i.e.*, $F_s$), crossover rate (*i.e.*, $C_r$), and total number of generations (*i.e.*, $n_g$) are defined as 20, 0.8, 0.7, and 400, respectively. The hyperparameters of the D-ICDE algorithm are mainly adopted from [21], [44], [45]. Given the above inputs to the SRO model and hyperparameters of the D-ICDE algorithm, a detailed breakdown of computational time of the implemented solution strategy is presented in Table III. The computations for this example are performed on a personal computer equipped with a six-core Intel Core i7 CPU with a core clock of 3.2 GHz and 16 GB of memory. The reported times are based on the average of all 400 generations. As the D-ICDE algorithm is ideal for parallel computing, the computations are performed on parallel on four cores. The total computational time for all 400 generations is approximately 1 hour which is reasonable for planning purposes.

TABLE III. COMPUTATIONAL TIME OF THE IMPLEMENTED SOLUTION STRATEGY FOR IEEE 33-BUS

| Computational time on a single core (s) | | | |
|---|---|---|---|
| Shock model | Self-healing model | Recovery model | Single generation of D-ICDE |
| 0.054 | 0.037 | 0.499 | 36.259 |

Fig. 3 presents the value of the objective function and the degree of constraint violation of individuals for all generations of the D-ICDE algorithm. Most of the individuals violate the constraints in the first few generations, after which they quickly become feasible. This result confirms the capability of ArATM in the D-ICDE algorithm in handling constrained optimization problems. It can be seen that the objective function value of all individuals converges to around $0.26 million for the last 100 generations, which indicates that none of the individuals in offspring populations can offer a more optimal solution.

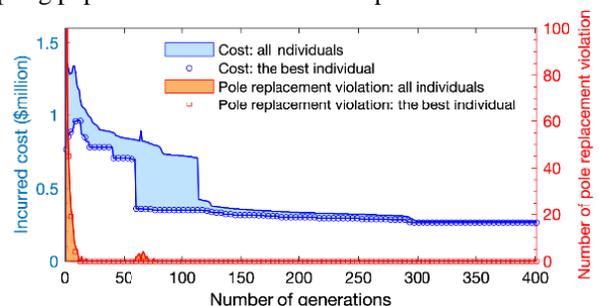

Fig. 3. Evolution of individuals in the implemented solution strategy for IEEE 33-bus system

Although the hardening budget is 50, the returned optimal hardening solution recommends replacing only 27 utility poles out of the 485 poles in the system before the impact of hurricane. These 27 poles are presented in Fig. 4.

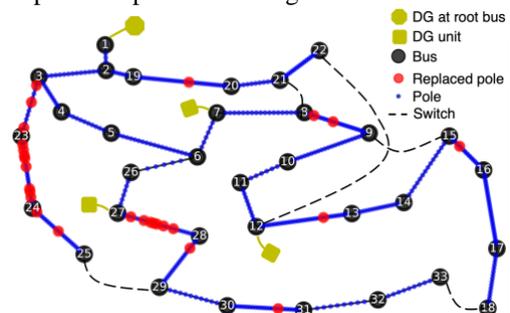

Fig. 4. Optimal hardening solution for IEEE 33-bus system

After hardening decisions are made, the shock model in the framework captures the impacts of a probabilistic hurricane event on the system by determining the failure of lines that can cause the expected worst-case scenario of damage. Based on the proposed model, the expected worst-case scenario is when lines (12,13), (20,21), and (28,29) fail with an annual failure probability of 0.141, 0.266, and 0.030 (after hardening), respectively. In the self-healing stage, the damaged system is reconfigured by closing switches at lines (9,15), (12,22), (18,33), and (25,29) to reduce the load shedding cost. According to (57), the expected time for repairing damaged lines (12,13), (20,21), and (28,29) by a single crew is 9.63, 9.83, and 9.12 hours, respectively. The recovery model recommends first repairing line (28,29), then line (20,21), and finally line (12,13). A summary of the recovery process is presented in Table IV. Following the optimal hardening strategy, immediate system reconfiguration, and recovery process, the incurred cost becomes $262,662.

To highlight the effectiveness of hardening measures for resilience enhancement, the result of the proposed SRO model is compared with another case where hardening actions are not applied to the system (*i.e.*, $x$ is set to $\mathbf{0}$). In this case, the shock model shows the failure of lines (12,13), (19,20), (3,23), and (27,28). The order of repairing failed lines is presented in Table IV. If the hardening actions are not applied to the system, the total incurred cost becomes $1,061,094, which indicates around 304% increase in the incurred cost compared to the case where optimal hardening actions are applied.

TABLE IV. OPTIMAL REPAIR CREW SCHEDULING

| Failed line $(i,j)$ | Time step of recovery (relative to $t_r$) (1-hour steps) | Assigned crew-hour | $\mathbb{E}[r_{ij}]$ (crew-hour) |
|---|---|---|---|
| With optimal hardening | | | |
| (12,13) | 11 | 10 | 9.63 |
| (20,21) | 7 | 10 | 9.83 |
| (28,29) | 4 | 10 | 9.12 |
| Without hardening | | | |
| (12,13) | 7 | 10 | 9.88 |
| (19,20) | 11 | 11 | 10.20 |
| (3,23) | 4 | 10 | 9.24 |
| (27,28) | 14 | 10 | 9.75 |

Fig. 5 shows changes in system performance as well as the cumulative incurred cost over time (relative to $t_e$) for both cases of the system with optimal hardening and without hardening. It can be observed that investing only $90,450 in hardening can save $798,432 per year. Based on the performance curves, optimal hardening not only increases minimum system performance to 63.12%, but also it expedites system recovery. It can also be observed that the topology reconfiguration after the event can significantly increase the system performance. More specifically, the performance of the hardened system reaches 63.12% after $t_e$ and it increases to 92.73% after reconfiguration in the self-healing stage. Finally, the system is fully recovered after 35 hours (relative to $t_e$). Whereas the performance of the system without hardening reaches 34.05% following the event and increases to 45.36% after self-healing stage with time to full recovery of 38 hours.

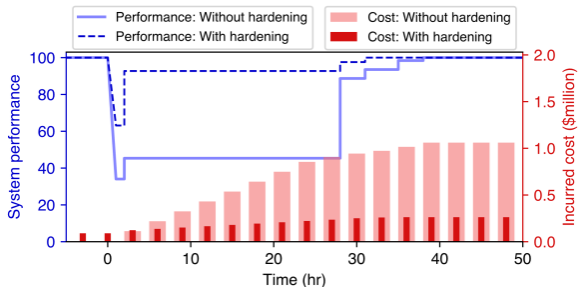

Fig. 5. Performance and incurred cost for reconfigurable 33-bus system

To properly quantify the performance of the system from the beginning of the hazard event to the end of the recovery time, the resilience of the system is calculated. The system resilience metric is defined as the normalized area under the system performance curve and is estimated as:

$$\mathcal{R}_{sys} = \frac{\int_{t_e}^{t_c} \mathcal{P}_{sys}(t) \, dt}{t_c - t_e} \qquad (58)$$

where $\mathcal{P}_{sys}(t)$ indicates the performance of the system which is computed as follows:

$$\mathcal{P}_{sys}(t) = 100 \left( 1 - \frac{\sum_{j \in \Omega_B} \rho_{j,t}^{ld} P_{j,t}^{ld}}{\sum_{j \in \Omega_B} P_{j,t}^{ld}} \right) \qquad (59)$$

According to (58), resilience of the system without hardening is 77.99%, while resilience of the hardened system reaches 96.51%. The results for the 33-bus system, including the incurred cost and resilience, are summarized in Table V.

To investigate the impacts of topology reconfiguration on system resilience and incurred cost, the performance and corresponding incurred cost for the system with optimal hardening and without hardening are evaluated with the assumption that the systems are not reconfigurable. The results of this analysis are presented in Fig. 6. According to this figure, when the systems are not reconfigurable, the incurred cost of the systems with optimal hardening and without hardening is $817,026 and $1,352,370, respectively, which indicates over 211% and 27% increase in the cost compared to the same but reconfigurable systems. When system reconfiguration is neglected in the analysis, resilience of the hardened system and the system without hardening becomes 85.01% and 71.89%, respectively. These values indicate over 13% and 8% decrease in the resilience of the non-reconfigurable hardened system and the system without hardening compared to their reconfigurable counterparts. Comparing system performance curves in Fig. 5 and Fig. 6 also reveals that co-optimization of repair and reconfiguration can significantly enhance the performance of the system during recovery. More specifically, according to Fig. 5, performance of the system without hardening is 45.36% once the recovery starts (*i.e.,* 24 hours after $t_e$); however, when the first failed line is repaired (*i.e.,* 28 hours after $t_e$), the performance reaches 88.70%, which is 43.34% increase in the performance. On the other hand, according to Fig. 6, performance of the system without hardening is 34.05% at the beginning of the recovery phase. In this case, the performance reaches 59.08% after repairing the first failed line, which indicates only 25.03% increase in the performance. It is obvious that system reconfiguration cannot affect the duration of repair; however, this important feature of active distribution systems can significantly restore power while the repair is in progress. This observation points to the significance of coordinating hardening and system operational measures for smart power distribution systems. To elaborate more, if corrective actions, such as topology reconfiguration and optimal recovery process, are not considered in simulating system operation during and post-event, the obtained incurred costs will not represent the actual costs. Therefore, hardening decisions are made based on incomplete or inaccurate estimates of the incurred cost, which can result in non-optimal hardening policies. It is worth noting that corrective actions that are taken during and after a hurricane event may not be necessarily optimal in the real world as those decisions rely on practical and real-time conditions, availability of information and the practices of utilities. However, this study assumes that smart distribution systems can be operated optimally in these conditions considering access to sensors and automated devices, such as fault indicators, line monitors, and remote-control switches, as well as advanced communication networks to collect and process real-time information.

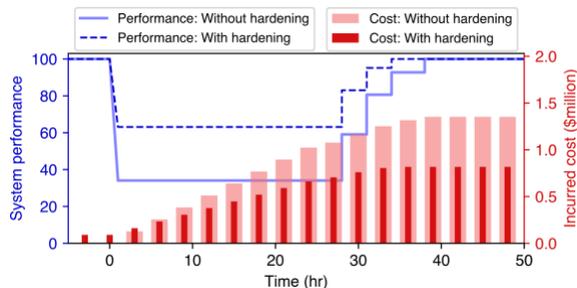

Fig. 6. Performance and incurred cost for non-reconfigurable 33-bus system

TABLE V. SUMMARY OF THE RESULTS OF THE SRO MODEL FOR IEEE 33-BUS

| Total incurred cost | | System resilience | |
|---|---|---|---|
| With hardening | Without hardening | With hardening | Without hardening |
| Reconfigurable system | | | |
| $262,662 | $1,061,094 | 96.51% | 77.99% |
| Non-reconfigurable system | | | |
| $817,026 | $1,352,370 | 85.01% | 71.89% |

In the last part of this subsection, we compare the performance of the proposed SRO model with a planning framework based on robust optimization similar to the one proposed by Ma et al. [8]. Major differences of this robust optimization model compared to the SRO model are: (a) hardening decisions are made for the entire line rather than individual poles; (b) hardening decisions are made based on a single predefined hurricane event; and (c) interactions of operational measures with hardening decisions are neglected. The results for this planning framework are summarized in Table VI for two cases: (1) the predefined hurricane event is Category-1 with a maximum wind speed of 74 mph; and (2) the predefined hurricane is Category-3 with a maximum wind speed of 111 mph. It should be noted that the same inputs that are listed at the beginning of this subsection are used for the robust optimization model in order to make a fair comparison. Furthermore, to make the results of the robust optimization model comparable with the results of the proposed SRO model, the reported cost and resilience in Table VI are calculated after evaluating outputs of the robust optimization model following steps 8-12 in Algorithm 1. More specifically, the hardening decision variable $x$ is set to hardening decisions based on the robust optimization model, and subsequently the proposed shock, self-healing, and recovery models are solved to obtain the incurred cost and performance of the system during and after the hurricane event. Then, the total incurred cost is computed by (2), (5), (47), and (48), and system resilience is calculated using (58). According to Table VI, using different hurricane events in the robust optimization model can result in different hardening decisions. This observation shows that existing planning frameworks based on robust optimization rely on the predefined hurricane event, which can increase the challenges of decision-making, as hurricanes have uncertain characteristics and assigning a predefined event does not capture those uncertainties. Comparing the presented results in Table V and Table VI, it can be seen that hardening decisions based on the robust optimization model result in significantly higher incurred cost and lower resilience compared to the case that the reconfigurable system is hardened based on the results of the SRO model. Non-optimal hardening solutions of the robust optimization model are due to the fact that decisions are made for entire lines, based on a single hurricane event, and without considering impacts of operational measures.

TABLE VI. RESULTS OF THE ROBUST OPTIMIZATION MODEL FOR IEEE 33-BUS

| Hardened lines | Total incurred cost | System resilience |
|---|---|---|
| Case 1: Category-1 hurricane | | |
| (3,23), (23,24) | $819,322 | 83.23% |
| Case 2: Category-3 hurricane | | |
| (2,19), (3,23) | $1,023,936 | 78.92% |

### E. Results for 118-bus system

All parameters used for analyzing the 118-bus distribution system are the same as the parameters used for the 33-bus system, except $B_H$. The hardening budget for the 118-bus system is set to 150 utility poles. A breakdown of computational time of the implemented solution strategy for 118-bus system is provided in Table VII. The computations for this example are performed on a computer equipped with a 12-core Intel Core i9-10920X CPU with a core clock of 3.50 GHz and 128 GB of memory. The reported times are based on the average of all 400 generations. The computations are performed in parallel on eight cores with the total computational time of less than 5 hours for all 400 generations.

TABLE VII. COMPUTATIONAL TIMES FOR 118-BUS SYSTEM

| Computational time on a single core (s) | | | |
|---|---|---|---|
| Shock model | Self-healing model | Recovery model | Single generation of D-ICDE |
| 0.501 | 0.313 | 4.647 | 354.549 |

Although the hardening budget is 150 for this case study, the optimal hardening solution indicates replacing only 110 utility poles out of the 1841 poles. These 110 poles are presented in Fig. 7. As can be seen in this figure, in most of the lines, only a few poles per line should be replaced to reach the optimal hardening strategy. However, in prior studies, hardening decisions were made for an entire line (*i.e.*, whether to harden the line or not). The results of this study confirm that making hardening decisions for a line yields sub-optimal solutions. For the optimal hardening strategy in Fig. 7, the shock model specifies failure of lines (22,23), (29,30), and (40,41) to cause the expected worst-case scenario of damage in the system. Then, the self-healing model based on reconfiguration recommends closing all switches, except switches at lines (9,40), (37,62), (88,75), (108,83), and (105,86). According to the recovery model, the optimal recovery process is achieved when the order of repair is to start with line (29,30), then line (40,41), and finally line (22,23). Following the optimal hardening strategy and operational resilience measures, the incurred cost of the 118-bus system becomes $2,084,992.

To evaluate impacts of hardening measures on system performance and incurred cost by the hurricane, we analyze the 118-bus system without hardening and compare the results with the previous case where hardening actions are applied to the system prior to the hurricane event occurrence. The results of this comparison are presented in Fig. 8. According to the results, the total incurred cost of the hardened system and the system without hardening are $2,084,992 and $3,312,384, respectively, which indicates that investing only $368,500 in hardening the system prior to the event can save as high as $1,227,392 annually. Furthermore, according to (58), resilience of the system without hardening is 88.53%, while resilience of the hardened system reaches 94.05%. These results highlight the importance of utility pole replacement as an effective approach for enhancing the resilience of distribution systems.

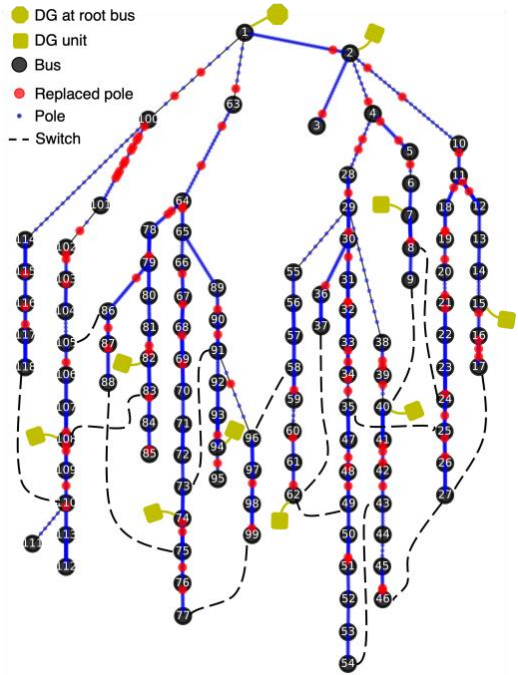

Fig. 7. Optimal hardening solution for 118-bus system

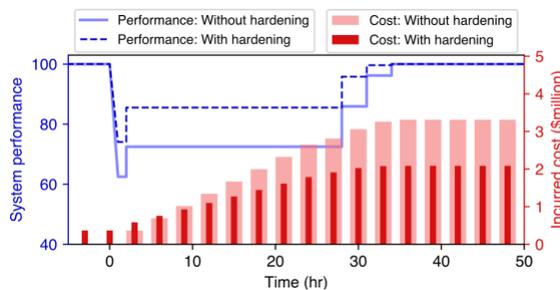

Fig. 8. Temporal pattern of performance and incurred cost for 118-bus system

## V. CONCLUSION

This paper develops a stochastic robust optimization (SRO) model for resilience planning of power distribution systems exposed to hurricanes. The primary objective of the SRO model is to find optimal hardening policies, while properly incorporating different sources of uncertainty and modeling impacts of operational measures on these policies. The SRO model consists of four stages, including hardening planning, operations to minimize incurred loss of function by hurricane shocks during the event, self-healing based on topology reconfiguration, and grid recovery in the aftermath of the event. The hardening stage provides optimal decisions for utility pole replacements to mitigate the likelihood of failure of these critical components. The shock stage models the impacts of probabilistic hurricane events on the system by failing lines given an uncertainty set. This stage captures the ability of the system to absorb disruptions by adjusting power flow in the damaged system. The bilevel stochastic robust mathematical programming model of this stage is transformed to its equivalent single-level problem. The self-healing stage is modeled as a single mixed-integer linear programming. This stage represents the ability of the system in autonomous service restoration after disruption. The recovery process is formulated as a single-level stochastic problem where a probabilistic model is developed to capture the uncertainty of the repair time of failed lines. In this stage, system operation decisions, including switch and power flow variables, and repair crew scheduling are determined interdependently. A solution strategy is designed based on the combination of the D-ICDE algorithm – a differential evolution method designed for constrained optimization problems with discrete variables – and a mixed-integer solver. In this resilience maximization model, hardening solutions are coupled with the shock and recovery stages through the developed uncertainty set and probabilistic repair time model, respectively. The resilience enhancement approach is applied to a modified version of IEEE 33-bus system integrated with four distributed generations and five tie switches as well as a 118-bus system with ten distributed generations and fifteen tie switches. These systems are mapped on the Harris County, TX, U.S. Results indicate that optimal hardening planning can decrease the expected incurred cost as high as 304% per year and increase the system resilience by over 18% compared to the case where no hardening is considered. It is also shown that topology reconfiguration feature of active distribution systems can reduce the expected incurred cost by over 211% annually and enhance the system resilience as high as 12%.

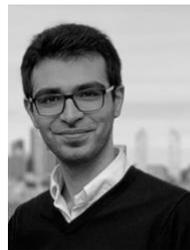

**Nariman L. Dehghani** received the B.S. in civil engineering from Iran University of Science and Technology, Tehran, Iran, in 2013, and the M.S. in earthquake engineering from Amirkabir University of Technology, Tehran, Iran, in 2016. He is currently pursuing the Ph.D. degree in structural engineering at The Ohio State University, OH, USA.

His primary research interests include sustainability and resilience quantification, reliability assessment of civil infrastructure systems, and optimal decision making for multi-state systems.

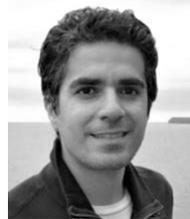

**Abdollah Shafieezadeh** received the B.S. and M.S. degrees in civil engineering from the University of Tehran, Tehran, Iran, in 2002 and 2006, respectively. He received a M.S. degree in structural engineering from the Utah State University, Logan, UT, USA, in 2008 and the Ph.D. degree in structural engineering with a minor in mathematics from the Georgia Institute of Technology, Atlanta, GA, USA, in 2011.

He is the Lichtenstein Associate Professor in the Department of Civil, Environmental and Geodetic Engineering at The Ohio State University, Columbus, OH, USA. His primary research interests are in uncertainty quantification using machine learning, and resilience quantification and optimal management of infrastructure systems.